\documentclass[twocolumn,showpacs,amsmath,amssymb,floatfix,aps]{revtex4}
\usepackage{graphicx}
\usepackage{dcolumn}
\usepackage{bm}
\begin{document}
\preprint{M. Yamada et al.}
\title{Interplay Between Time-Temperature-Transformation and\\ 
the Liquid-Liquid Phase Transition in Water}
\author{Masako~Yamada$^1$, Stefano~Mossa$^{1,2}$,
H.~Eugene~Stanley$^1$, and Francesco~Sciortino$^2$}
\affiliation{
$^1$ Center for Polymer Studies and Department of Physics,
Boston University, Boston, Massachusetts 02215\\
$^2$
Dipartimento di Fisica, INFM UdR  and INFM Center for
Statistical Mechanics and Complexity,
Universit\`{a} di Roma ``La Sapienza'', Piazzale Aldo
Moro 2, I-00185, Roma, Italy
}
\date{\today}
\begin{abstract}
We study the TIP5P water model proposed by Mahoney
and Jorgensen, which is closer to real water 
than previously-proposed classical pairwise additive potentials. 
We simulate  the model in a wide range of deeply supercooled states and 
find {\em (i)} the existence of a non-monotonic ``nose-shaped'' 
temperature of maximum density  line and a non-reentrant spinodal, 
{\em (ii)} the presence of a low temperature phase transition,
{\em (iii)} the free evolution of bulk water to ice, and
{\em (iv)} the time-temperature-transformation curves at different densities.
\end{abstract}
\maketitle
Much effort has been invested in exploring the overall
phase diagram of water and the connection among its liquid, supercooled
and glassy states~\cite{Debenedetti96,Smith99,Mishima94}, 
with particular interest in understanding the origin of 
the striking anomalies at low
temperatures, such as the $T$-dependence of the 
isothermal compressibility $K_T$,
the constant pressure specific heat $C_P$, 
and the thermal expansivity $\alpha_P$.

The ``stability limit conjecture'' attributes the
increase of the response functions upon supercooling to a continuous
re-tracing spinodal line bounding the superheated,
supercooled and stretched (negative pressure) metastable
states~\cite{Speedy82}. This line at its minimum
intersects the temperature of maximum density (TMD) curve
tangentially.  More recently, a different hypothesis has been
developed, for which the spinodal does not
re-enter into the positive pressure region, but rather the anomalies are
attributed to a critical point below the homogeneous nucleation
line~\cite{Poole92}.  The TMD line, which is negatively sloped at
positive pressures, becomes positively sloped at sufficiently negative
pressures and does not intersect the spinodal.
A line of first order phase transitions --- interpreted as the liquid state
analog of the line separating low and high density amorphous
glassy phases~\cite{Poole92,Mishima94} ---
develops from this critical point.

Simulations of supercooled metastable states are possible
because the structural relaxation time at the temperatures of interest is
several orders of magnitude shorter than the crystallization time. It is
difficult, but not impossible~\cite{kusalik95}, 
to observe crystallization in simulations of molecular
models~\cite{Huitema00} because homogeneous nucleation rarely occurs on
the time scales reachable by present day computers. 
Bulk water simulations
have been crystallized by applying a homogeneous electric
field~\cite{Svishchev94} or  placing liquid water in contact with
pre-existing ice~\cite{Borzsak99,Clancy95}, 
but spontaneous crystallization of deeply
supercooled model water has not been observed in simulations.

In contrast, experimental measurements of metastable liquid states are
strongly affected by homogeneous nucleation. The
nucleation and growth of ice particles from aqueous solution has been
extensively studied, and the  ``nose-shaped''
time-temperature-transformation (TTT) curves have been
measured~\cite{Debenedetti96,MacFarlane83,Kresin91}. 
The non-monotonic relation between
crystallization rate and supercooling depth results from the
competition between the thermodynamic driving force for nucleation and the
kinetics of growth~\cite{Debenedetti96}. Crystallization hinders
direct experimental investigation of pure metastable
liquid water below the homogeneous nucleation line; only indirect
measurements can be made by
studying the metastable melting lines of
ices~\cite{Mishima94}. 

This work attempts to unify the phenomena connected with the existence 
of a liquid-liquid phase transition and 
homogeneous nucleation in a single molecular dynamics simulation study. 
We simulate a system of $N=343$ molecules interacting with the
TIP5P potential~\cite{Mahoney00}.
TIP5P is a five-site, rigid, non-polarizable water model,
not unlike the ST2 model~\cite{Stillinger74}. The TIP5P 
potential accurately reproduces
the density anomaly at 1 atm and exhibits excellent structural
properties when compared with experimental
data~\cite{Mahoney00,Head-Gordon}. The
TMD shows the correct pressure dependence, shifting 
to lower temperatures as pressure is increased. Under ambient conditions, 
the diffusion constant is close to the
experimental value, with reasonable temperature
and pressure dependence away from ambient conditions~\cite{Mahoney00}. 
\begin{figure} 
\centering
\includegraphics[width=.47\textwidth]{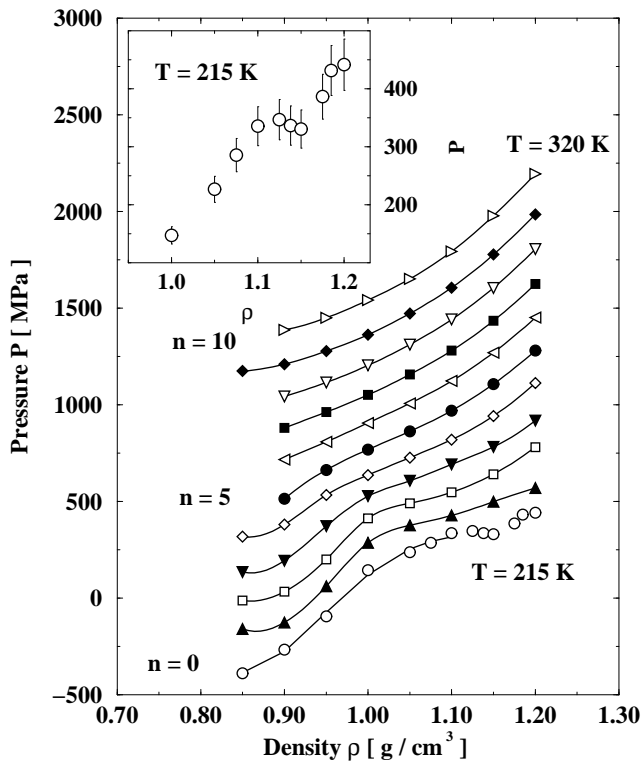}
\caption{
\label{fig:isotherms}
Dependence on density of the pressure at all temperatures investigated
($T=215,220,230,240,250,260,270,280,290,300,320$ K, from bottom to top). 
Each curve has been shifted by $n\times 150$ MPa to avoid overlaps.
An inflection appears as T is decreased, transforming into a
``flat'' coexistence region at $T=215$ K, indicating the presence of a
liquid-liquid transition.
Inset: A detailed view of the $T=215$ K isotherm.
}
\end{figure}

We perform equilibration runs  at constant $T$ 
(Berendsen's thermostat), while we perform
production runs  in the microcanonical (NVE) ensemble.
After  thermalization at  $T =320$~K we set
the thermostat temperature  to
the temperature of interest. We let  
the system  evolve for a time longer than the structural relaxation 
time $\tau_\alpha$, defined as 
 the time at which  $F_s(Q_0,\tau_\alpha)=1/e$, where 
$F_s(Q_0,t)$ is  the self-intermediate
scattering function evaluated at $Q_0=18$ nm$^{-1}$,  the location of 
the first peak of the static structure factor.
In the time  $\tau_\alpha$, each molecule diffuses on average  a distance
of the order of the nearest neighbor distance.
We use the final configuration of the equilibration run
to start a production run 
of length greater than several $\tau_\alpha$ and then analyze
the calculated trajectory.
We check that no drift in any of the 
studied quantities and no crystallization occurs during the production
run.
\begin{figure}[t]
\centering
\includegraphics[width=.47\textwidth]{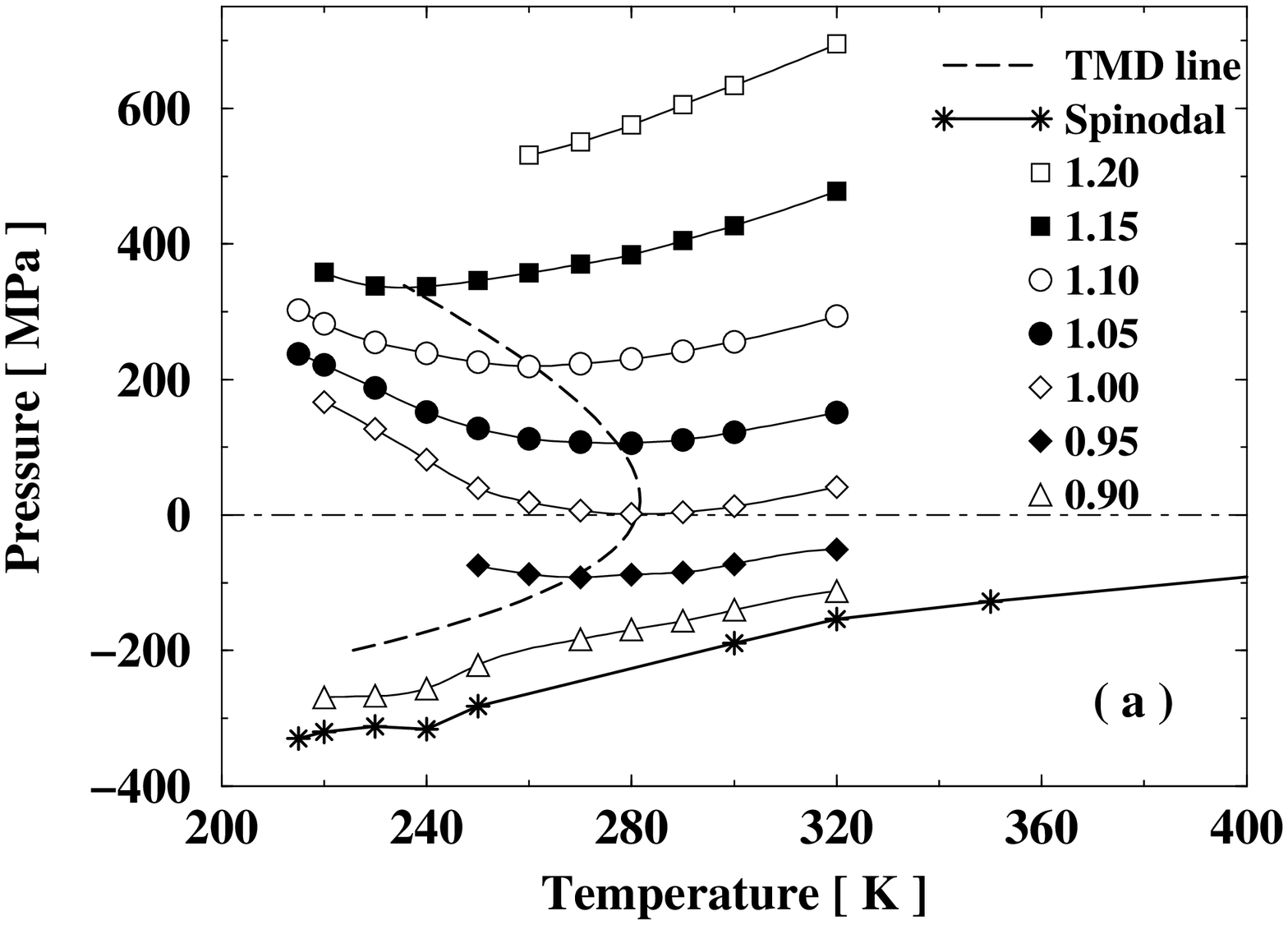}

\includegraphics[width=.47\textwidth]{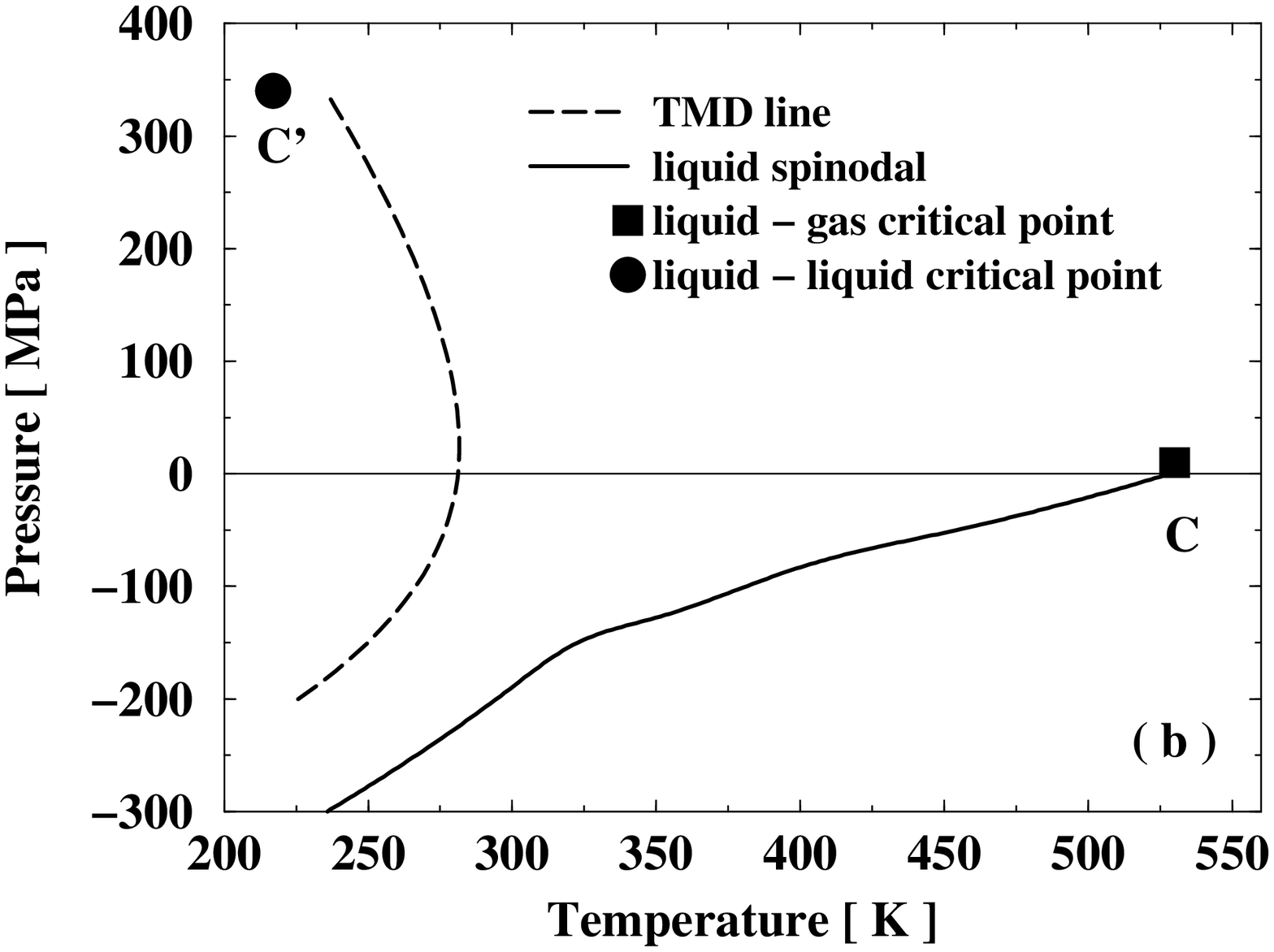} 
\caption{
\label{fig:phase_diagram} 
{\em (a)} Pressure along seven isochores;
the minima correspond to the temperature of
maximum density line (dashed line). Note the ``nose'' of the TMD
line at $T=4$ C. Stars denote
the liquid spinodal line, which is not reentrant, and terminates at the 
liquid-gas critical point. {\em (b)} The full phase diagram of TIP5P
water. The liquid-gas critical point $C$ is indicated by  the filled
square\cite{highT} and liquid-liquid critical point $C'$ by the filled
circle.
} 
\end{figure} 

In Fig.~\ref{fig:isotherms} we show results for pressure along
isotherms. At lower temperatures an 
inflection develops, which becomes a ``flat'' isotherm at the
lowest temperature, $T=215$ K.
The presence of a flat region indicates that a
phase separation takes place, and we estimate the critical temperature
$T_{C'}=217 \pm 3$ K, the critical pressure  $P_{C'}=340\pm$ 20 MPa, and
the critical density  $\rho_{C'}=1.13 \pm 0.04$ g/cm$^3$.
\begin{figure}
\includegraphics[width=.45\textwidth]{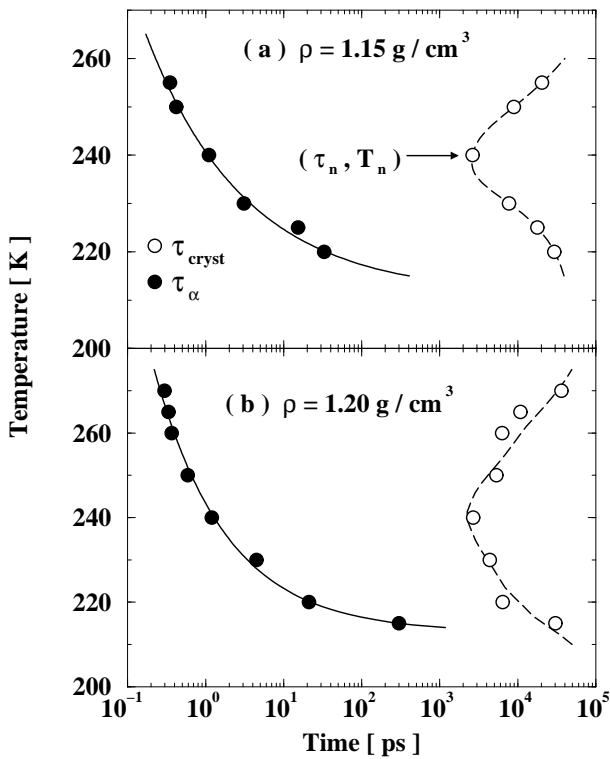} 
\caption{
\label{fig:crystallization_time} 
Average crystallization time (open circles) as a function of temperature at 
the two densities {\em (a)} $1.15$ g/cm$^3$ and
{\em (b)} $1.20$ g/cm$^3$. A well-defined nose shape is visible, as
measured
for water solutions~\cite{MacFarlane83}. We also show the structural
relaxation times $\tau_\alpha$ as calculated from the self-intermediate
scattering function $F_s(Q,t)$ (closed circles). Shown as solid lines are 
the MCT power law fits with $T_c=211$ K, $\gamma=2.9$ 
for $\rho = 1.15$ g/cm$^3$, and $T_c=213$ K, $\gamma=2.13$ 
for $\rho = 1.20$ g/cm$^3$.   
The interplay between these two curves is discussed in the text.
} 
\end{figure} 

In Fig.~\ref{fig:phase_diagram}(a) we plot the pressure along isochores.  
The curves show minima as a function of temperature; the locus of the
minima is the TMD line, since $(\partial P/\partial T)_V=\alpha_P /K_T$.
It can be seen that the pressure exhibits a minimum if the density
passes through a maximum ($\alpha=0$).
It is clear that, as in the case of
ST2 water, TIP5P water has a TMD that changes slope from negative to
positive as P decreases. Notably, the point of crossover between the two
behaviors is located  at ambient pressure,  $T\approx 4 $ C , and
$\rho \approx 1$ g/cm$^{3}$. 

We also plot the spinodal line. We calculate the points 
on the spinodal line fitting the isotherms (for $T\ge 300 K$) of
Fig.~\ref{fig:isotherms} to the form
$P(T,\rho)=P_s(T)+A\left[\rho-\rho_s(T)\right]^2$, where $P_s(T)$  
and $\rho_s(T)$ denote
the pressure and density of the spinodal line.
This functional form is the mean field prediction for 
$P(\rho)$ close to a spinodal line. 
For $T\le 250 K$, we calculate $P_s(T)$ by estimating the location of the 
minimum of $P(\rho)$.
The results in Fig.\ref{fig:phase_diagram} show that the liquid spinodal
line is not reentrant and does not intersect the TMD line~\cite{highT}. 

A supercooled liquid  is metastable with respect to the 
crystal, so it is driven to crystallize~\cite{Debenedetti96}. 
However, crystallization of model water has not been found in 
simulations because the homogeneous nucleation time
far exceeds the CPU time. However, for TIP5P, 
crystallization times
lie within a time window accessible to present-day
simulations, and we observe 
crystallization at densities $\rho=1.15$ g/cm$^3$, and  $1.20$ g/cm$^3$ 
for a wide range of temperatures (Fig.~\ref{fig:crystallization_time}).
\begin{figure}
\centering
\hspace{2cm}
\includegraphics[width=.7\textwidth]{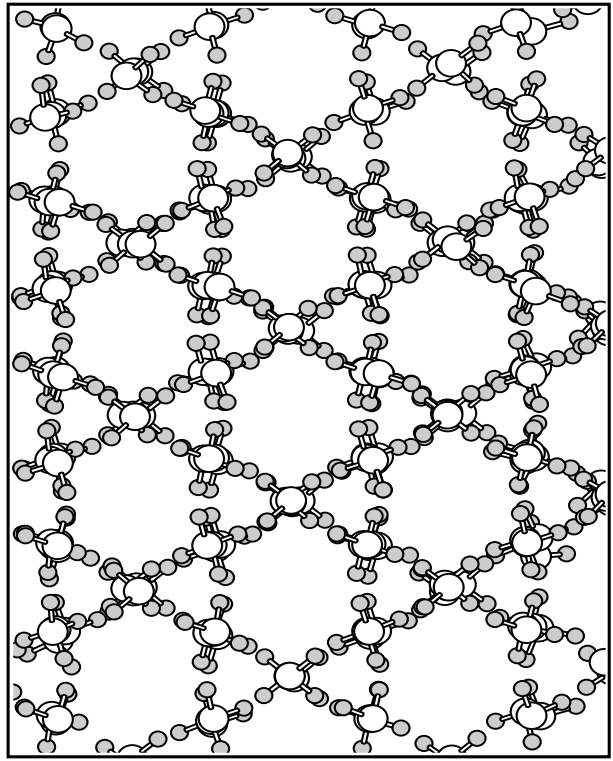}

\vspace{-0.5cm}
\caption{
\label{fig:snapshot} 
For $\rho$ = 1.20 g / cm$^3$, the energy minimum configuration, 
viewed along the $c$-axis, of a crystal formed at $T = 270$ K.
} 
\end{figure} 

To quantify the crystallization process, 
we analyze four independent configurations thermalized at
temperature $T=320$ K and instantaneously quenched to the temperature of
interest. We monitor the potential energy as well as the time evolution of
the structure factor
$S_{\vec{Q}}(t) \equiv 
\langle \rho_{\vec{Q}}(t)\rho^*_{\vec{Q}}(t)\rangle/N$ at all 
wave vector $\vec{Q}$ values, ranging from the smallest value $2\pi/L$
allowed by the side $L$ of the simulation box up to
$50$ nm$^{-1}$. The oxygen density fluctuation
$\rho_{\vec{Q}}(t) \equiv \sum_{i=1}^N\exp{ (i {\vec Q}\cdot {\vec
r}_i)}$, where ${\vec r}_i$ 
is the oxygen coordinate of molecule $i$.
The onset of crystallization coincides with the
occurrence of (i) a sudden drop in potential energy and (ii) a
sharp increase in the density fluctuations at one or more wave vector
values. When crystallization occurs, the value of 
$S_{\vec{Q}}(t)$ jumps from ${\cal O}$(1) in the liquid to ${\cal O}$(N).

Defining the crystallization time is somewhat arbitrary because of the
stochasticity which accompanies 
the onset of crystallization and the definition of the critical
nucleus.
We define $\tau_{cryst}$ as the time at which {\em any}
density fluctuation  $S_{\vec{Q}}(t)$
grows above a  threshold value $S^*= 15$ and remains
continuously above the threshold for a time exceeding $t^*= 40$ ps. This
threshold prevents
transient density fluctuations from being attributed to crystallization. We 
also perform calculations for other definitions of $S^*$ and $t^*$,
but the above values are sufficient to unambiguously identify the
onset of crystallization without requiring excessive simulation time. 
Fig.~\ref{fig:crystallization_time} shows
the crystallization times $\tau_{cryst}$, averaged over the 
four independent runs,  for two different densities and
for a broad range of $T$. The resulting TTT curve shows 
a characteristic ``nose'' shape, arising from the competition between 
two effects, the thermodynamic driving force for nucleation and the kinetics of 
growth~\cite{Debenedetti96,kusalik95}. As
temperature is lowered, both the thermodynamic driving force and the
relaxation time increase, and it becomes more difficult for
particles
to diffuse to the energetically-preferred crystalline configuration.
For both densities, $\rho=1.15$, and  $1.20$ g/cm$^3$, 
the $T$ at which nucleation is fastest is
around $240$ K. At this $T$, the onset of crystallization 
requires about $3$ ns. At the lowest studied
$T$, the crystallization time has grown to $30$ ns.

Fig.~\ref{fig:crystallization_time} also shows the
relaxation times $\tau_\alpha$. The $T$ dependence of $\tau_\alpha$ 
can be described by 
a power law
$\tau_\alpha\propto(T-T_c)^\gamma$, in agreement with 
the prediction of  mode coupling theory \cite{leshouches}.
Since the relation $\tau_{\alpha}\ll \tau_{cryst}$
holds at each temperature, including in the deeply supercooled region, 
``equilibrium'' studies of metastable water can be achieved before
nucleation takes place. 
The liquid  can be connected to the deeply supercooled state  
via ``equilibrium'' metastable states if we choose a quench rate 
larger than the  critical cooling rate ${\cal R}_c
\equiv  (T_m-T_n)/\tau_n$, where $T_m$ is the melting $T$ and 
$T_n$ and $\tau_n$ locate the nose in the TTT curve \cite{Debenedetti96}.
For TIP5P, ${\cal R}_c \approx 10^{10}$ K/s 
at the two studied densities. 
For $\rho=1.10$ g/cm$^3$ and $T=240$ K, we observe only one (out of four) 
crystallization event within a time of 70 ns.
For densities smaller than $\rho=1.10$ g/cm$^3$, we observe no
crystallization events within a time of 60 ns and hence we can
only estimate that ${\cal R}_c$ is smaller than 
$10^{9}$ K/s (the experimental value   for water at 
ambient pressure is  ${\cal R}_c \approx 10^{7}$ K/s~\cite{uhlmann72}).

In  Fig.~\ref{fig:snapshot} we show a typical crystal configuration. 
The crystal structure, after energy minimization at
constant volume, is a proton-ordered structure similar to ice-B,
first observed by Baez and Clancy\cite{Clancy95}. Ice-B is a 
variant of the ice IX structure, which is the proton-ordered 
form of ice III. The density of 
ice IX and ice III is in fact $1.16$ g/cm$^{3}$, close to our value.

We have shown that the liquid-liquid phase separation can be observed
in metastable equilibrium 
{\it (i)} if the  cooling rate is faster than ${\cal R}_c$, and
{\it (ii)} if the observation time is shorter than the crystallization time at the
critical point. While both such conditions can be realized in numerical
simulations --- as shown here --- 
they cannot be met in experiments.  
Our simulations also show that a continuity of states between liquid 
and glassy phases of water exists~\cite{Hallbrucker89}. Liquid states 
below the crystallization temperature can be accessed 
provided the cooling rate exceeds ${\cal R}_c$. 

We acknowledge useful discussions with D.~R.~Baker, G.~Franzese, W.~Kob,
E.~La Nave, M.~Marquez and C.~Rebbi. We acknowledge support from the NSF
Grant CHE-0096892. M.~Y. acknowledges support from NSF Grant GER-9452651
as a Graduate Research Trainee at the Boston University Center for
Computational Science. F.~S. acknowledges support from MURST COFIN 2000
and INFM Iniziativa Calcolo Parallelo.

%
%
%

%
%

\begin{thebibliography}{99}
%
\bibitem{Debenedetti96}
P.~G.~Debenedetti, {\em Metastable Liquids} 
(Princeton University Press, Princeton, 1996).
%
\bibitem{Smith99}
R.~S.~Smith and B.~D.~Kay, Nature (London) {\bf 398}, 788 (1999).
%
\bibitem{Mishima94}
O.~Mishima, J. Chem. Phys. {\bf 100}, 5910 (1994); 
Phys. Rev. Lett. {\bf 85}, 334 (2000);
O.~Mishima and H.~E.~Stanley, Nature (London) {\bf 392}, 164 (1998); 
{\it ibid.} {\bf 396}, 329 (1998). 
%
\bibitem{Speedy82}
R.~J.~Speedy, J. Chem. Phys. {\bf 86}, 982 (1982).
%
\bibitem{Poole92}
P.~H.~Poole {\it et al.},
Nature (London) {\bf 360}, 324 (1992); Phys. Rev. E {\bf 48}, 3799 (1993); 
Phys. Rev. E {\bf 48}, 4605 (1993); Phys. Rev. E {\bf 55}, 3799 (1997).
%
\bibitem{kusalik95}
I.~M.~Svishchev and P.~G.~Kusalik, Phys. Rev. Lett. {\bf 75}, 3289 (1995).
%
\bibitem{Huitema00}
For the case of atomic systems see, e.g.
H.~E.~A.~Huitema, J.~P.~van der Eerden, J.~J.~M.~Janssen, and H.~Human,
Phys. Rev. B {\bf 62}, 14690 (2000).
%
\bibitem{Svishchev94}
I.~M.~Svishchev and P.~G.~Kusalik, Phys. Rev. Lett. {\bf 73}, 975 (1994).
%
\bibitem{Borzsak99}
I.~Borzs\'ak and P.~T.~Cummings, Chem. Phys. Lett. {\bf 300}, 359 (1999).
%
\bibitem{Clancy95}
L.~A.~Baez and P.~Clancy, J. Chem. Phys. {\bf 103}, 9744 (1995).
%
\bibitem{MacFarlane83}
D.~R.~MacFarlane, R.~K.~Kadiyala, and C.~A.~Angell, 
J. Chem. Phys. {\bf 79}, 3921 (1983).
%
\bibitem{Kresin91}
M.~Kresin and Ch. K\"{o}rber, J. Chem. Phys. {\bf 95}, 3921 (1991).
%
\bibitem{Mahoney00}
M.~W.~Mahoney and W.~L.~Jorgensen, J. Chem. Phys. {\bf 112}, 8910
(2000);  {\it ibid} {\bf 114}, 363 (2001). 
%
\bibitem{Stillinger74}
F.~H.~Stillinger and A.~Rahman, J. Chem. Phys. {\bf 60}, 1545 (1974).
%
\bibitem{Head-Gordon}
J.~M.~Sorenson, G.~Hura, R.~M.~Glaeser, and T.~Head-Gordon, J. Chem. Phys.
{\bf 113}, 9149 (2000).
%
\bibitem{highT}
We also calculated the TIP5P isotherms at high $T$ to provide an
estimate of the location of the   liquid-gas critical point,
the terminus of the liquid spinodal line. We find
$T_{C}\sim$ 530 $\pm 10$ K, $\rho_{C}\sim$ $0.33 \pm 0.08 $ g/cm$^3$,
$P_{C}\sim$ 10 $\pm 2 $ MPa, in rough agreement with the experimental
values.
%
\bibitem{leshouches}
W. G\"otze, in {\it Liquids, Freezing, and Glass Transition, 
Proc. Les Houches}, edited by J.~P.~Hansen, D.~Levesque, and J.~Zinn-Justin
(North-Holland, Amsterdam, 1991).
%
\bibitem{uhlmann72}
D.~R.~Uhlmann, J. Non-cryst. Solids {\bf 7}, 337 (1972).
%
\bibitem{Hallbrucker89}
A.~Hallbrucker {\it et al.}, J. Phys. Chem. {\bf 93}, 4986 (1989).
%
%
\end{thebibliography}
\end{document}